\documentclass[useAMS]{mn2e}
\usepackage{amsmath,amssymb,graphicx,soul,color,longtable,float}
\usepackage{subfigure}

\title[Polarization in active dwarfs]{Broad-band linear polarization in late-type active dwarfs}
\author[Patel et al.]{Manoj K. Patel$^1$\thanks{patelmanoj79@gmail.com}, Jeewan C. Pandey$^2$\thanks{jeewan@aries.res.in}, Subhajeet Karmakar$^2$, D. C. Srivastava$^1$, \newauthor and Igor S. Savanov$^3$\\
$^{1}$Department of Physics, Deen Dayal Upadhyay Gorakhpur University, Gorakhpur - 273009, India\\
$^{2}$Aryabhatta Research Institute of Observational Sciences (ARIES), Nainital - 263 001, India\\
$^{3}$Institute of Astronomy, Russian Academy of Sciences, ul. Pyatnitskaya 48, Moscow, 119017 Russia}

\begin{document}

\date{} 

\pagerange{\pageref{firstpage}--\pageref{lastpage}} \pubyear{2015}

\maketitle     \label{firstpage}

\begin{abstract}
We present recent polarimetric results of magnetically active late-type  dwarfs. The polarization in these stars  is found to be wavelength dependent, decreasing towards the longer wavelength. The average values of degree of polarization in these active dwarfs are found to be $0.16\pm0.01$, $0.080\pm0.006$, $0.056\pm0.004$ and $0.042\pm0.003$ per cent in $B$, $V$, $R$, and $I$ bands, respectively. Present results indicate that polarization in the majority of active dwarfs are primarily due to sum of the polarization by magnetic intensification and scattering. However, supplementary sources of the polarization are also found to  be  present in some  active stars.  The correlations between the degree of polarization and various activity parameters like Rossby number, chromospheric activity indicator and coronal activity indicator are found to be stronger in $B$ band and weaker in $I$ band.
\end{abstract}


\section{Introduction}
The inhomogeneity due to the magnetic areas (star spots or plages) and circumstellar gas or dust envelopes on the surface of magnetically active late-type dwarfs may produce a broad-band linear polarization (BLP; Tinbergen \& Zwaan 1981). The polarization of the integrated stellar light may change along with the stellar activity phenomenon (Huovelin et al. 1988, 1989). The connection of magnetic field regions with BLP was first shown by Dollfus (1958) and Leroy (1962), and observed by Piirola (1977). The sources of linear polarization in late type active dwarfs are still unclear. However, the most popular mechanisms those have been suggested by many studies are magnetic intensification and scattering due to circumstellar material. The BLP  in cool active stars increases towards shorter wavelengths. Such wavelength dependent polarization also supports the models for  magnetic and scattering origins of polarization (Huovelin et. al. 1985; Leroy et al. 1989). The number of saturated magnetically sensitive absorption lines increases towards shorter wavelengths and represents the  characteristic of the observed wavelength dependence polarization (Kemp \& Wolstencroft 1974; Calamai, Landi Degl'Innocenti \& Landi Degl'Innocenti 1975; Tinbergen \& Zwaan 1981). But a similar effect could be produced by Rayleigh scattering in the optically thin part of the stellar atmosphere. Magnetic intensification in the saturated Zeeman sensitive lines will arise from stellar spots and plages (Leroy 1962; Mullan \& Bell 1976; Huovelin \& Saar 1991; Saar \& Huovelin 1993), while the scattering could take place in optically thin discs or other inhomogeneities above the stellar surface (Brown \& McLean 1977; Brown, McLean \& Emslie 1978; Pfeiffer 1979; Piirola \& Vilhu 1982).

 Measurements of BLP have been made for numerous late-type active dwarfs in the past. Huovelin et al. (1985) have made polarimetric studies in $U$, $B$, $V$, $R$, and $I$ bands of a sample of 13 solar-type stars and have found that the intrinsic polarization decreases towards longer wavelengths only for the stars HD 20630, HD 20998, HD 126053, and HD 206860. Leroy \& Le Borgne (1989) conducted a survey of 30 solar-type stars and found no evidence of intrinsic polarization. Apart from these surveys, many other solar-type stars are observed polarimetrically and the polarization is explained  either by magnetic origin or scattering of thin circumstellar material or sum of both.  Some studies show that the degree of linear polarization depends on the photospheric, chromospheric, and coronal activity levels of the star and follows the short period changes in activity caused by rotational modulation, suggesting a magnetic origin (see Tinbergen \& Zwaan 1981; Huovelin et. al. 1988, 1989; Alekseev 2000, 2003).

Evidence for a connection between linear polarization and magnetic activity indicators is lacking because on one hand the variation, as well as the mean degree of polarization of late-type dwarfs are very small, and  on the other hand sizes of samples studied are small.  Therefore, we have observed a sample of 50 active late type dwarfs and supplemented it with another sample of 18 active dwarfs, those are available in the literature (Huovelin et al. 1985; Alekseev 2000, 2003; Golovin et al. 2012; Patel et al. 2013). The criteria of selection of stars in the sample were (i) the $V$-band magnitude range of $6.0 - 13.0$ mag, and (ii) the ($B-V$) colour range of $0.5 - 1.5$ mag. Since these late-type active stars exhibit a very small value polarization, therefore, the faint limit of V-mag was set as $13$ mag for our telescope and instrument. Allowing the observations fainter than this limit show large error bars due to photon statistics. However, the lower limit of V-mag was set $6.0$ mag as going brighter than this limit the CCD  saturates even in a  very small exposure.  Basic parameters of these stars are summarized in Table \ref{tab:object}.
 
 The paper is organized as follows. We  describe observations and data reduction in the next section, while analysis and results are given in Section \ref{sec:Results}.  In Section \ref{sec:Correlation} we provide the correlation of linear polarization with activity parameters, and in the last section, we present discussion and conclusions.
                                                             
\begin{table*}
\centering
\small
\caption{Basic parameters of the stars in the sample.}
\label{tab:object}
\begin{tabular}{|c|l|c|c|c|c|c|c|c|c|}
\hline
S.No& ~~~Object&  Sp. type&    $V$     & ($B-V$) & Period & D  &log($R_{0}$)$^\star$  & log($R_{X}$)$^\star$ & log(R$^\prime_{HK}$)$^\star$   \\
    &          &          &   (mag)   &  Mag  & (d) & (pc)   &           &              &         \\
\hline                                                                  
\multicolumn{10}{|c|}{Present observed sample}               \\
1  & PW And    & K2V     &  8.852   & 0.936 & 01.76  & 032.40 & -1.103    & -3.232       &...      \\
2  & BK Psc    & K5V     &  10.57   & 1.160 & 02.16  & 035.92 & -1.049    & -3.000       &...      \\ 
3  & BD -04 234& G0V     &  9.961   & 0.550 & ...    & ...    &  ...      & ...          &...      \\
4  & V1221 Tau & G6V/K3V &  9.490   & 0.533 & 00.55  & 121.01 & -1.010    & -4.005       &...      \\           
5  & V988 Tau  & K2V     &  9.370   & 0.950 & 09.90  & 060.14 & -0.357    & -4.632       & -4.394  \\ 
6  & V526 Aur  & G0V     &  10.83   & 0.601 & 02.85  &...     &...        & ...          & ...     \\
7  & V538 Aur  & K0V     &  6.210   & 0.840 & 11.00  & 012.28 & -0.273    & -4.742       & -4.254  \\           
8  & V848 Mon  & K2V     &  8.950   & 0.960 & 19.98  & 030.94 & -0.054    & -3.820       & -4.419  \\ 
9  & V850 Mon  & K3V     &  9.310   & 0.940 & 09.90  & 054.27 & -0.355    &  ...         & -4.313  \\ 
10 & V429 Gem  & K5V     &  9.930   & 1.150 & 02.78  & 030.00 & -0.939    & -3.420       & -4.047  \\
11 & V867 Mon  & K2V     &  8.144   & 0.883 & 04.76  & 024.53 & -0.656    & -4.531       & -4.218  \\ 
12 & BD+231799 & K0V     &  9.980   & 1.090 & 20.03  & 034.40 & -0.376    & -2.855       &  ...    \\
13 & HO Cnc    & K5V     &  9.508   & 1.022 & 05.21  & 043.86 & -0.648    &  ...         & -4.257  \\ 
14 & HP Cnc    & K0V     &  9.060   & 0.746 & 11.14  & 030.76 & -0.367    & -3.999       & -3.480  \\   
15 & V402 Hya  & K0/K1V  &  8.980   & 0.770 & 00.41  & 031.16 & -0.900    & -3.050       &  ...    \\
16 & V405 Hya  & K2V     &  8.760   & 0.990 & 08.64  & 028.28 & -0.423    & -4.697       & -4.190  \\
17 & GT Leo    & K0V     &  8.881   & 0.902 & 10.92  & 049.13 & -0.303    &  ...         & -4.390  \\ 
18 & V417 Hya  & K2V     &  8.124   & 0.913 & 10.74  & 024.02 & -0.313    & -5.084       & -4.199  \\ 
19 & LR Hya    & K1.5V   &  8.020   & 0.855 & 06.87  & 034.22 & -0.485    & -4.746       & -4.190  \\
20 & V418 Hya  & K2V     &  8.750   & 0.850 & ...    & 032.21 &  ...      & -5.066       & -4.249  \\ 
21 & AB Crt    & K3V     &  9.040   & 0.990 & 10.33  & 029.14 & -0.346    &  ...         & -4.333  \\ 
22 & HL Leo    & G2V     &  7.400   & 0.850 & 37.17  & 222.94 &  0.250    & -4.095       & -3.977  \\ 
23 & GQ Leo    & K5Ve    &  10.90   & 1.020 & 04.45  & 050.00 & -0.717    & -2.897       &  ...    \\ 
24 & PQ Vir    & K0V     &  9.130   & 0.890 & ...    & 047.94 &  ...      &  ...         & -4.208  \\   
25 & MY Uma    & K0V     &  9.570   & 0.990 & 11.43  & 045.49 & -0.302    &  ...         & -4.300  \\ 
26 & PR Vir    & K3V     &  9.490   & 0.950 & 17.16  & 041.46 & -0.118    &  ...         & -4.303  \\ 
27 & FZ Leo    & G5V     &  8.430   & 0.560 & 00.91  & 102.07 & -0.880    & -3.253       &  ...    \\ 
28 & LV Com    & K2V     &  9.160   & 0.930 & 05.50  & 040.69 & -0.609    &  ...         & -4.433  \\ 
29 & DO CVn    & K0V     &  8.540   & 0.960 & 08.37  & 026.95 & -0.432    & -4.083       & -4.219  \\ 
30 & LX Com    & K1V     &  9.100   & 0.890 & 07.74  & 037.47 & -0.448    & -4.603       & -4.151  \\ 
31 & PX Vir    & K1V     &  7.690   & 0.847 & 06.47  & 021.68 & -0.507    & -4.288       & -4.108  \\
32 &GY Boo     & K0V     &  8.900   & 0.970 & 09.52  & 033.25 & -0.378    & -4.081       & -4.233  \\ 
33 & GZ Boo    & K2V     &  8.870   & 0.900 & 07.52  & 042.32 & -0.464    & -4.559       & -4.203  \\ 
34 & HO Boo    & K2V     &  7.960   & 0.850 & 93.00  & 025.15 &  0.649    & -4.858       & -4.341  \\ 
35 & FN Boo    & K6V     &  8.090   & 1.430 & 79.82  & 030.43 & -1.400    &  ...         &  ...    \\ 
36 & V1022 Her & K7V     &  11.71   & 1.504 & 00.81  & 030.00 & -1.525    & -2.950       &  ...    \\ 
37 & V1658 Aql & K0V     &  9.350   & 1.110 & 18.76  & 409.00 &  ...      & -2.930       &  ...    \\ 
38 & V1659 Aql & K0V     &  8.870   & 1.080 & 05.18  & 339.00 &  ...      &  ...         &  ...    \\ 
39 & V401 Vul  & K3V     &  10.83   & 1.030 & 02.16  & 064.03 & -1.033    & -3.045       &  ...    \\ 
40 & V402 Peg  & K2V     &  7.730   & 0.911 & 04.51  & 019.75 & -0.690    & -5.416       & -4.606  \\
41 & OT Peg    & K0V     &  9.740   & 0.810 & ...    & 074.04 &  ...      & -3.065       &  ...    \\
42 & V383 Lac  & K1V     &  8.580   & 0.840 & 02.47  & 034.00 & -0.921    & -3.155       & -3.773  \\
43 & HD 218782 & K2V     &  9.870   & 0.890 & 00.92  & ...    & -1.373    & -3.110       &         \\ 
\hline
\multicolumn{10}{|c|}{The sample taken from literature} \\
44  & HD 1835   & G3V    &  06.38 & 0.670   & 07.70  & 20.86  & -0.213    & -4.622       & -4.263  \\
45  & HD 20630  & G5V    &  04.85 & 0.660   & 09.40  & 09.14  & -0.164    & -4.659       & -4.208  \\
46  & HD 25998  & F7V    &  05.50 & 0.470   & 02.60  & 20.99  & -0.027    & -4.430       & -4.255  \\
47  & V1147 Tau & K5V    &  11.07 & 1.220   & 01.49  & 45.07  & -1.229    & -2.980       & ...     \\  
48  & FR Cnc    & K5V    &  10.24 & 1.110   & 00.83  & 34.07  & -1.460    & -3.289       & -3.480  \\
49  & LQ Hya    & K0V    &  07.83 & 0.920   & 01.70  & 18.62  & -1.116    & -3.081       & -3.728  \\
50  & HD 114378 & F5V    &  05.22 & 0.450   & 03.02  & 17.82  &  0.087    & -4.963       & -5.216  \\
51  & HD 114710 & G0V    &  04.25 & 0.580   & 12.35  & 09.13  &  0.204    & -5.702       & -4.597  \\
52  & HD 115383 & G0V    &  05.22 & 0.590   & 03.33  & 17.56  & -0.391    & -4.451       & -4.219  \\
53  & HD 126053 & G1V    &  06.30 & 0.600   & 21.10  & 17.19  &  0.290    &  ...         & -4.704  \\
54  & EK Dra    & G5V    &  07.61 & 0.640   & 02.67  & 34.12  & -0.631    & -3.553       & -3.887  \\
55  & HD 142373 & F9V    &  04.62 & 0.560   & 15.00  & 15.89  &  0.353    &  ...         & -5.210  \\
56  & MS Ser    & K2V    &  08.21 & 1.000   & 09.59  & 77.14  & -0.380    & -3.564       & -3.760  \\
57  & HD 154417 & F9V    &  06.01 & 0.580   & 07.80  & 20.66  &  0.004    & -4.946       & -4.359  \\
58  & HD 182101 & F6V    &  06.36 & 0.400   & 02.00  & 37.55  &  0.223    & -5.139       & -4.496  \\
59  & HD 187013 & F7V    &  04.99 & 0.470   & 06.40  & 21.23  &  0.347    & -5.471       & -4.639  \\
60  & HD 201091 & K5V    &  05.21 & 1.180   & 37.90  & 03.48  &  0.200    & -5.256       & -4.448  \\
61  & HD 206860 & G0V    &  05.95 & 0.580   & 04.70  & 17.88  & -0.230    & -4.462       & -4.194  \\ 
\hline
\end{tabular}
~~\\
D is distance\\
$^\star$-Parameters derived in this paper.
\end{table*}

\section{Observations and Data Reduction}
\label{sec:Object}
Polarimetric observations of stars in the sample were obtained during the years 2010 - 2014 using ARIES Imaging Polarimeter (AIMPOL; Rautela et al. 2004 ) mounted at the Cassegrain focus of the 104-cm Sampurnanand telescope (ST) of the Aryabhatta Research Institute of Observational Sciences, Nainital, India, coupled with a $1024 \times 1024$ CCD camera. AIMPOL consists of a half-wave plate (HWP) modulator and a Wollaston prism beam-splitter. Each pixel of the CCD corresponds to 1.73 arcsec and the field of view is $\sim$ 8 arcmin in diameter. The read-out noise and the gain of the CCD are 7.0$e^{-}$ and 11.98e$^{-}$ ADU$^{-1}$, respectively. For linear polarimetry, HWP is rotated with an interval of  $22^{\circ}.5$  between exposures. Thus, one polarization measurement was obtained from every four exposures at the HWP position of $0^{\circ}$, $22^{\circ}.5$, $45^{\circ}$ and $67^{\circ}.5$. The exposure time at each position of the HWP was same; ranging from 5 to 300 s depending on the filter used.  The full width at half-maximum of the stellar image varied from 2 to 3 pixels. Due to the absence of a grid in AIMPOL, we manually checked for any overlap of ordinary and extraordinary images of the sources.

 Fluxes of ordinary ($I_{o}$) and extraordinary ($I_{e}$) beams for all the observed sources with a good signal-to-noise ratio were extracted by standard aperture photometry after bias subtraction using the {\sc iraf}\footnote{iraf.net} package. We have calculated the  ratio R($\alpha$) as 

\begin{equation}
R(\alpha) = \frac{I_{e}(\alpha)-I_{o}(\alpha)} {I_{e}(\alpha)+I_{o}(\alpha)} =  P cos(2\theta - 4\alpha)
\end{equation}

\noindent
where $P$ is the fraction of the total linearly polarized light and $\theta$ is the polarization angle of the plane of polarization. Here, $\alpha$ is the position of the fast axis of the HWP at $0^\circ$, $22^{\circ}.5$, $45^{\circ}$ and $67^{\circ}.5$ corresponding to the four normalized Stokes parameters, respectively, $q [R(0^\circ)]$, $u [R(22^{\circ}.5)]$, $q_{1} [R(45^{\circ})]$ and $u_{1} [R(67^{\circ}.5)]$. The detailed descriptions about the AIMPOL, data reduction, calculations of polarization, and position angle are given in Rautela et al. (2004).

In order to correct the measurements for null polarization (or instrumental polarization) and the zero-point polarization angle, we observed several polarized and unpolarized standard stars in each night of observations. Stars HD 236633, HD 204827, HD 19820, HD 25443, HD 154445, and BD+$59^{\circ}~389$ were observed as standard polarized stars, while stars $\beta$ UMa, $\theta$ UMa, HD 14069, BD+$33^{\circ}~2642$, and GD 319 were observed as standard unpolarized stars. The observed degree of polarization [$P (\%)$] and polarization angle [$\theta (^\circ)$] for the polarized standards are similar to the standard values as given by Schmidt et al. (1992). These measurements show that the instrumental polarization of ST is below 0.1 per cent in all passbands, which was found to be constant over the past few years (see Soam et al. 2014). The instrumental polarization was then applied to all measurements. The values of degree of polarization and polarization position angle of active dwarfs as observed at different epochs are given in Table \ref{tab:pol}.  The data of seven stars were of bad quality due to poor sky conditions. Therefore, our final sample consists only 43 stars.
 
\begin{table*}
{\scriptsize
\caption[Degree of polarization........]{Degree of polarization and polarization angle of late-type active dwarfs in $B$, $V$, $R$, and $I$ bands.}\label{tab:pol}
\begin{tabular}{lccccccccc}
\hline 
\multicolumn{1}{l}{Object} &
\multicolumn{1}{l}{Mean epoch$^\star$} & 
\multicolumn{2}{c}{$B$-Filter} & 
\multicolumn{2}{c}{$V$-Filter} & 
\multicolumn{2}{c}{$R$-Filter} & 
\multicolumn{2}{c}{$I$-Filter} \\        
       & (HJD)       &P (per cent)& $\theta(^\circ)$ &P (per cent) & $\theta(^\circ)$& P (per cent) & $\theta(^\circ)$& P (per cent) & $\theta(^\circ)$ \\ [0.5ex] \hline
PW And    & 2456948.1136  & $0.216\pm0.043$ & $40\pm 5 $ & $0.099\pm0.041$ & $93\pm  6 $& $0.059\pm0.085$ & $157\pm15$ & $0.034\pm0.013$ & $25\pm3$ \\
          & 2456949.0839  & $0.211\pm0.032$ & $86\pm 5 $ & $0.079\pm0.019$ & $75\pm  3 $& $0.098\pm0.072$ & $91\pm11 $ & $0.053\pm0.086$ & $103\pm16$ \\
BK Psc    & 2455854.2780  & $0.250\pm0.076$ & $ 81\pm 6$ & $0.082\pm0.043$ & $ 77\pm 7$ & $0.065\pm0.027$ & $122\pm 6$ & $0.033\pm0.011$ & $101\pm 3$ \\
BD-04234  & 2456982.1425  & $0.473\pm0.099$ & $ 46\pm 5$ & $0.095\pm0.053$ & $136\pm 8$ & $0.079\pm0.072$ & $98\pm35 $ & $0.039\pm0.024$ & $174\pm 5$ \\
          & 2456982.5918  & $0.359\pm0.027$ & $ 33\pm 3$ & $0.093\pm0.080$ & $113\pm 21$& $0.054\pm0.038$ & $132\pm8 $ & $0.047\pm0.034$ & $71\pm 7$ \\
V1221 Tau & 2455887.2871  & $0.151\pm0.029$ & $  9\pm 3$ & $0.066\pm0.058$ & $ 80\pm17$ & $0.040\pm0.064$ & $ 33\pm13$ & $0.028\pm0.051$ & $ 82\pm20$ \\  
V988 Tau  & 2455886.3006  & $0.143\pm0.018$ & $ 11\pm 2$ & $0.072\pm0.043$ & $ 47\pm17$ & $0.044\pm0.028$ & $ 49\pm 6$ & $0.042\pm0.076$ & $161\pm15$ \\
          & 2456003.1265  & $0.142\pm0.074$ & $ 15\pm15$ & $0.064\pm0.034$ & $149\pm 6$ & $0.050\pm0.070$ & $130\pm13$ & $0.048\pm0.076$ & $ 19\pm15$ \\
V526 Aur  & 2455559.3357  & $0.413\pm0.068$ & $155\pm 4$ & $0.474\pm0.042$ & $179\pm 3$ & $0.405\pm0.091$ & $170\pm 5$ & $0.362\pm0.027$ & $176\pm 2$ \\
          & 2455601.1475  & $0.678\pm0.021$ & $168\pm 2$ & $0.669\pm0.044$ & $177\pm 2$ & $0.606\pm0.035$ & $178\pm 1$ & $0.610\pm0.025$ & $175\pm 1$ \\
V538 Aur  & 2456978.7681  & $0.302\pm0.026$ & $28\pm 3 $ & $0.090\pm0.042$ & $86\pm13  $& $0.058\pm0.078$ & $22\pm15 $ & $0.044\pm0.010$ & $133\pm3$ \\
V848 Mon  & 2455886.3394  & $0.151\pm0.010$ & $ 20\pm 2$ & $0.065\pm0.023$ & $ 91\pm10$ & $0.061\pm0.026$ & $158\pm55$ & $0.037\pm0.070$ & $122\pm15$ \\
          & 2456002.1748  & $0.162\pm0.012$ & $ 64\pm 2$ & $0.098\pm0.141$ & $ 75\pm20$ & $0.052\pm0.048$ & $ 19\pm 9$ & $0.050\pm0.030$ & $129\pm 4$ \\
V850 Mon  & 2455887.3991  & $0.197\pm0.040$ & $ 11\pm 4$ & $0.086\pm0.045$ & $ 98\pm15$ & $0.050\pm0.010$ & $ 46\pm 3$ & $0.032\pm0.013$ & $110\pm 3$ \\ 
          & 2456036.0566  & $0.208\pm0.030$ & $173\pm 4$ & $0.085\pm0.057$ & $143\pm 9$ & $0.062\pm0.034$ & $ 82\pm 6$ & $0.038\pm0.031$ & $ 17\pm 7$ \\
V429 Gem  & 2456978.8045  & $0.268\pm0.028$ & $60\pm9  $ & $0.113\pm0.077$ & $12\pm20  $& $0.083\pm0.062$ & $173\pm10$ & $0.070\pm0.035$ & $173\pm10$ \\
          & 2456980.3207  & $0.261\pm0.058$ & $24\pm6  $ & $0.106\pm0.053$ & $179\pm7  $& $0.087\pm0.034$ & $61\pm13 $ & $0.064\pm0.021$ & $180\pm3$  \\
V867 Mon  & 2455886.3711  & $0.117\pm0.088$ & $169\pm12$ & $0.051\pm0.039$ & $145\pm 7$ & $0.032\pm0.081$ & $141\pm18$ & $0.056\pm0.070$ & $122\pm13$ \\
          & 2456015.1162  & $0.160\pm0.026$ & $  4\pm 3$ & $0.064\pm0.026$ & $ 35\pm 5$ & $0.042\pm0.010$ & $132\pm 2$ & $0.045\pm0.052$ & $100\pm10$ \\
BD+231799 & 2455603.2027  & $0.165\pm0.020$ & $147\pm 2$ & $0.065\pm0.021$ & $ 14\pm 4$ & $0.044\pm0.068$ & $158\pm13$ & $0.042\pm0.076$ & $ 41\pm15$ \\
          & 2456002.2215  & $0.141\pm0.010$ & $ 56\pm 2$ & $0.078\pm0.059$ & $ 24\pm10$ & $0.041\pm0.057$ & $139\pm13$ & $0.034\pm0.010$ & $ 97\pm 2$ \\
HO Cnc    & 2455886.4332  & $0.264\pm0.020$ & $ 33\pm 2$ & $0.102\pm0.028$ & $ 19\pm 4$ & $0.071\pm0.035$ & $ 22\pm14$ & $0.045\pm0.027$ & $ 56\pm 5$ \\
          & 2455972.2458  & $0.221\pm0.014$ & $165\pm 3$ & $0.103\pm0.064$ & $ 27\pm18$ & $0.058\pm0.022$ & $140\pm10$ & $0.044\pm0.039$ & $ 92\pm26$ \\
HP Cnc    & 2455887.4411  & $0.135\pm0.016$ & $ 36\pm 2$ & $0.057\pm0.021$ & $ 38\pm 4$ & $0.038\pm0.014$ & $ 47\pm 3$ & $0.030\pm0.054$ & $ 86\pm12$ \\            & 2455972.2841  & $0.152\pm0.074$ & $ 79\pm14$ & $0.084\pm0.059$ & $ 98\pm 9$ & $0.064\pm0.024$ & $149\pm11$ & $0.034\pm0.076$ & $ 90\pm16$ \\
V402 Hya  & 2456036.0990  & $0.265\pm0.093$ & $ 16\pm 7$ & $0.090\pm0.052$ & $ 57\pm17$ & $0.068\pm0.052$ & $ 11\pm22$ & $0.044\pm0.062$ & $ 91\pm12$ \\
V405 Hya  & 2455559.4573  & $0.114\pm0.017$ & $113\pm 2$ & $0.067\pm0.010$ & $ 92\pm 3$ & $0.025\pm0.042$ & $104\pm48$ & $0.044\pm0.037$ & $127\pm 7$ \\
          & 2455602.3067  & $0.151\pm0.067$ & $114\pm 8$ & $0.056\pm0.021$ & $ 94\pm 4$ & $0.048\pm0.023$ & $ 98\pm 5$ & $0.039\pm0.046$ & $ 53\pm10$ \\
GT Leo    & 2455588.3479  & $0.117\pm0.011$ & $174\pm 3$ & $0.056\pm0.066$ & $ 85\pm34$ & $0.037\pm0.041$ & $ 95\pm10$ & $0.025\pm0.031$ & $101\pm 5$ \\
          & 2455593.3627  & $0.130\pm0.082$ & $103\pm10$ & $0.071\pm0.045$ & $146\pm 8$ & $0.032\pm0.037$ & $ 80\pm 7$ & $0.044\pm0.005$ & $  2\pm 2$ \\
V417 Hya  & 2455956.3962  & $0.131\pm0.014$ & $ 39\pm 3$ & $0.069\pm0.059$ & $ 22\pm10$ & $0.038\pm0.069$ & $133\pm14$ & $0.031\pm0.016$ & $ 39\pm 3$ \\
          & 2455986.3307  & $0.124\pm0.053$ & $178\pm12$ & $0.074\pm0.018$ & $ 75\pm 7$ & $0.055\pm0.037$ & $101\pm19$ & $0.038\pm0.030$ & $124\pm 6$ \\
LR Hya    & 2456984.0450  & $0.181\pm0.033$ & $94 \pm 5$ & $0.125\pm0.011$ & $100\pm 2 $& $0.081\pm0.037$ & $84\pm6  $ & $0.032\pm0.015$ & $84\pm13$ \\
          & 2456984.5619  & $0.135\pm0.028$ & $19\pm 15$ & $0.089\pm0.023$ & $17\pm  4 $& $0.060\pm0.003$ & $75\pm3  $ & $0.042\pm0.024$ & $135\pm5$ \\
V418 Hya  & 2456002.2661  & $0.120\pm0.016$ & $ 26\pm 2$ & $0.059\pm0.032$ & $ 69\pm 6$ & $0.038\pm0.010$ & $174\pm 2$ & $0.022\pm0.045$ & $ 60\pm11$ \\
          & 2456035.0968  & $0.112\pm0.048$ & $ 97\pm 7$ & $0.064\pm0.021$ & $110\pm14$ & $0.074\pm0.045$ & $109\pm 9$ & $0.035\pm0.030$ & $ 94\pm 7$ \\
AB Crt    & 2456001.2376  & $0.211\pm0.010$ & $ 24\pm 2$ & $0.092\pm0.010$ & $ 32\pm 2$ & $0.056\pm0.019$ & $ 71\pm10$ & $0.042\pm0.011$ & $ 21\pm 2$ \\
          & 2456015.2655  & $0.247\pm0.032$ & $ 85\pm 4$ & $0.088\pm0.030$ & $ 22\pm 5$ & $0.057\pm0.010$ & $176\pm 2$ & $0.040\pm0.024$ & $ 61\pm 5$ \\
HL Leo    & 2455578.7580  & $0.082\pm0.066$ & $ 49\pm10$ & $0.035\pm0.020$ & $ 60\pm16$ & $0.035\pm0.036$ & $163\pm30$ & $0.024\pm0.030$ & $ 57\pm10$ \\
          & 2455603.3665  & $0.095\pm0.020$ & $ 70\pm 6$ & $0.032\pm0.070$ & $153\pm15$ & $0.030\pm0.100$ & $ 18\pm22$ & $0.026\pm0.052$ & $ 67\pm11$ \\
GQ Leo    & 2456036.1684  & $0.248\pm0.049$ & $ 81\pm 4$ & $0.092\pm0.059$ & $ 72\pm 3$ & $0.065\pm0.031$ & $ 62\pm 5$ & $0.044\pm0.037$ & $ 78\pm 7$ \\
          & 2456092.1181  & $0.222\pm0.026$ & $ 57\pm 3$ & $0.078\pm0.040$ & $ 77\pm10$ & $0.086\pm0.010$ & $117\pm 2$ & $0.042\pm0.031$ & $ 38\pm 8$ \\
PQ Vir    & 2455603.4791  & $0.144\pm0.027$ & $ 69\pm 3$ & $0.071\pm0.020$ & $108\pm 8$ & $0.059\pm0.010$ & $ 88\pm 2$ & $0.025\pm0.064$ & $ 99\pm15$ \\
          & 2455955.3870  & $0.137\pm0.072$ & $ 46\pm 9$ & $0.055\pm0.011$ & $ 42\pm 2$ & $0.048\pm0.023$ & $140\pm14$ & $0.029\pm0.016$ & $ 58\pm 4$ \\
MY UMa    & 2456001.2803  & $0.140\pm0.011$ & $ 31\pm 2$ & $0.061\pm0.034$ & $139\pm 6$ & $0.056\pm0.015$ & $ 95\pm 3$ & $0.033\pm0.040$ & $116\pm 9$ \\
          & 2456013.3025  & $0.151\pm0.086$ & $ 87\pm16$ & $0.078\pm0.044$ & $115\pm 7$ & $0.057\pm0.012$ & $ 68\pm 6$ & $0.034\pm0.010$ & $154\pm 2$ \\
PR Vir    & 2455955.4279  & $0.214\pm0.081$ & $ 99\pm11$ & $0.081\pm0.016$ & $ 44\pm 6$ & $0.055\pm0.033$ & $150\pm 6$ & $0.064\pm0.025$ & $104\pm11$ \\
          & 2455986.3740  & $0.209\pm0.016$ & $109\pm 2$ & $0.076\pm0.048$ & $ 14\pm 8$ & $0.058\pm0.012$ & $136\pm 3$ & $0.061\pm0.050$ & $118\pm 9$ \\
FZ Leo    & 2456019.3231  & $0.112\pm0.024$ & $ 25\pm 3$ & $0.054\pm0.016$ & $ 56\pm 3$ & $0.039\pm0.029$ & $120\pm10$ & $0.029\pm0.059$ & $ 42\pm 9$ \\
          & 2456035.1380  & $0.103\pm0.092$ & $ 41\pm13$ & $0.053\pm0.036$ & $ 67\pm 7$ & $0.036\pm0.025$ & $106\pm20$ & $0.024\pm0.031$ & $123\pm 7$ \\
LV Com    & 2455588.4741  & $0.150\pm0.028$ & $110\pm 6$ & $0.087\pm0.023$ & $ 91\pm 4$ & $0.061\pm0.047$ & $  2\pm22$ & $0.066\pm0.069$ & $122\pm12$ \\
          & 2455602.4095  & $0.126\pm0.052$ & $ 37\pm12$ & $0.053\pm0.017$ & $ 77\pm 3$ & $0.047\pm0.056$ & $160\pm11$ & $0.058\pm0.057$ & $144\pm 8$ \\
DO CVn    & 2456001.3195  & $0.133\pm0.060$ & $ 40\pm 8$ & $0.068\pm0.056$ & $ 51\pm 9$ & $0.047\pm0.052$ & $105\pm10$ & $0.029\pm0.010$ & $114\pm 2$ \\
          & 2456004.3271  & $0.151\pm0.020$ & $ 34\pm 4$ & $0.060\pm0.018$ & $ 34\pm 8$ & $0.050\pm0.015$ & $149\pm 9$ & $0.025\pm0.025$ & $ 99\pm 6$ \\
LX Com    & 2455602.4754  & $0.157\pm0.061$ & $124\pm 7$ & $0.087\pm0.013$ & $143\pm 2$ & $0.044\pm0.068$ & $105\pm14$ & $0.021\pm0.042$ & $134\pm12$ \\
          & 2455956.4363  & $0.137\pm0.014$ & $ 44\pm 2$ & $0.070\pm0.022$ & $166\pm 9$ & $0.061\pm0.018$ & $ 10\pm 3$ & $0.026\pm0.038$ & $165\pm43$ \\
PX Vir    & 2456984.5779  & $0.213\pm0.091$ & $62\pm 8 $ & $0.135\pm0.035$ & $60\pm  4 $& $0.076\pm0.055$ & $48\pm21 $ & $0.065\pm0.020$ & $33\pm4$ \\
GY Boo    & 2456002.4543  & $0.120\pm0.010$ & $ 44\pm 2$ & $0.045\pm0.022$ & $112\pm14$ & $0.031\pm0.002$ & $102\pm 2$ & $0.026\pm0.032$ & $135\pm 7$ \\
          & 2456015.4263  & $0.124\pm0.019$ & $ 42\pm 4$ & $0.049\pm0.050$ & $ 98\pm10$ & $0.050\pm0.060$ & $138\pm12$ & $0.047\pm0.042$ & $110\pm 8$ \\
GZ Boo    & 2456016.4076  & $0.207\pm0.010$ & $ 82\pm 2$ & $0.080\pm0.042$ & $ 46\pm16$ & $0.066\pm0.012$ & $  6\pm 2$ & $0.033\pm0.075$ & $148\pm65$ \\
          & 2456036.3709  & $0.204\pm0.073$ & $ 60\pm 7$ & $0.086\pm0.038$ & $ 42\pm 6$ & $0.059\pm0.021$ & $ 92\pm 4$ & $0.038\pm0.088$ & $128\pm18$ \\
HO Boo    & 2455986.4308  & $0.134\pm0.095$ & $148\pm20$ & $0.055\pm0.014$ & $144\pm 3$ & $0.036\pm0.057$ & $ 71\pm12$ & $0.021\pm0.018$ & $115\pm 4$ \\
          & 2456002.4285  & $0.117\pm0.051$ & $ 49\pm 7$ & $0.049\pm0.048$ & $ 43\pm 9$ & $0.036\pm0.013$ & $106\pm 3$ & $0.020\pm0.080$ & $143\pm20$ \\
FN Boo    & 2455294.2500  & $0.250\pm0.024$ & $ 95\pm 6$ & $0.087\pm0.011$ & $103\pm 3$ & $0.054\pm0.011$ & $ 99\pm 2$ & $0.021\pm0.037$ & $ 85\pm10$ \\
          & 2455337.3210  & $0.234\pm0.010$ & $132\pm 2$ & $0.082\pm0.020$ & $127\pm 2$ & $0.027\pm0.098$ & $147\pm23$ & $0.032\pm0.087$ & $ 64\pm20$ \\
V1022 Her & 2456091.2666  & $0.302\pm0.086$ & $ 49\pm 6$ & $0.122\pm0.025$ & $ 16\pm 3$ & $0.080\pm0.015$ & $173\pm 5$ & $0.057\pm0.049$ & $119\pm25$ \\ 
          & 2456094.1142  & $0.300\pm0.071$ & $ 54\pm 5$ & $0.142\pm0.052$ & $ 67\pm 6$ & $0.078\pm0.034$ & $110\pm14$ & $0.050\pm0.051$ & $100\pm 9$ \\
V1658 Aql & 2455854.0749  & $0.443\pm0.052$ & $ 75\pm 4$ & $0.418\pm0.058$ & $ 82\pm 3$ & $0.350\pm0.081$ & $ 89\pm 5$ & $0.370\pm0.021$ & $ 88\pm 2$ \\
          & 2455886.0550  & $0.659\pm0.101$ & $ 98\pm 4$ & $0.762\pm0.118$ & $ 99\pm 4$ & $0.639\pm0.146$ & $ 84\pm 7$ & $0.627\pm0.131$ & $ 81\pm13$ \\
V1659 Aql & 2455854.0749  & $0.550\pm0.047$ & $ 82\pm 2$ & $0.575\pm0.041$ & $ 76\pm 2$ & $0.535\pm0.023$ & $ 96\pm 2$ & $0.530\pm0.080$ & $ 88\pm 4$ \\
          & 2455886.0550  & $0.555\pm0.012$ & $ 78\pm 1$ & $0.550\pm0.032$ & $ 89\pm 2$ & $0.550\pm0.064$ & $103\pm 3$ & $0.498\pm0.054$ & $ 91\pm 3$ \\
V401 Vul  & 2455890.0710  & $0.234\pm0.054$ & $  9\pm 5$ & $0.129\pm0.106$ & $ 75\pm13$ & $0.065\pm0.048$ & $  4\pm21$ & $0.030\pm0.011$ & $121\pm 3$ \\
V402 Peg  & 2456948.0589  & $0.102\pm0.036$ & $26\pm 5 $ & $0.040\pm0.078$ & $82\pm 16 $& $0.030\pm0.065$ & $122\pm8 $ & $0.022\pm0.039$ & $18\pm33$ \\
          & 2456949.0326  & $0.115\pm0.024$ & $68\pm 10$ & $0.030\pm0.064$ & $163\pm14 $& $0.024\pm0.053$ & $103\pm35$ & $0.014\pm0.045$ & $164\pm24$ \\
OT Peg    & 2455527.1293  & $0.161\pm0.039$ & $ 78\pm 4$ & $0.055\pm0.025$ & $ 80\pm 4$ & $0.054\pm0.010$ & $ 67\pm 2$ & $0.031\pm0.004$ & $ 37\pm 2$ \\
          & 2455559.0736  & $0.152\pm0.076$ & $ 96\pm 7$ & $0.078\pm0.010$ & $121\pm 2$ & $0.039\pm0.054$ & $111\pm11$ & $0.026\pm0.031$ & $103\pm 8$ \\
V383 Lac  & 2456949.0679  & $0.116\pm0.059$ & $33\pm 8 $ & $0.066\pm0.027$ & $176\pm 5 $& $0.033\pm0.012$ & $28\pm3  $ & $0.015\pm0.042$ & $85\pm10$ \\
          & 2456948.0992  & $0.128\pm0.039$ & $168\pm9 $ & $0.088\pm0.058$ & $105\pm 9 $& $0.060\pm0.017$ & $70\pm13 $ & $0.028\pm0.065$ & $80\pm10$ \\
HD 218782 & 2455886.1614  & $0.265\pm0.042$ & $ 27\pm 3$ & $0.166\pm0.015$ & $ 49\pm 2$ & $0.078\pm0.080$ & $101\pm13$ & $0.045\pm0.078$ & $ 78\pm15$ \\
\hline
\end{tabular}

$^\star$ mean epoch of $B$, $V$, $R$ and $I$ observations
}
\end{table*}

\section{Analysis and Results}
\label{sec:Results}

\begin{figure}
\centering
\includegraphics[width=90mm]{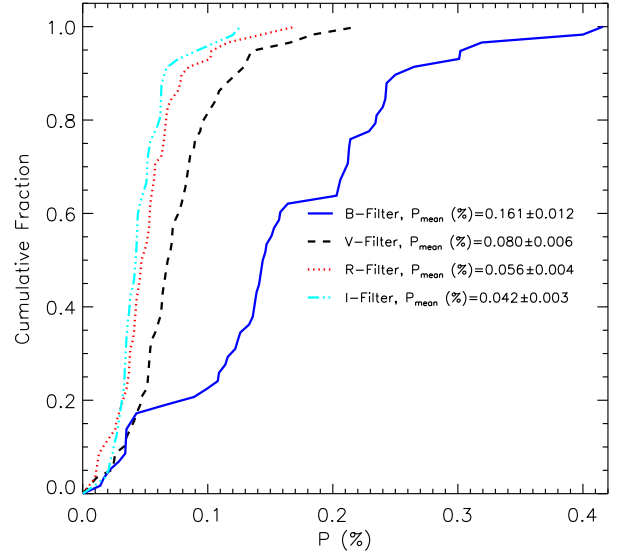}
\caption{Cumulative distribution of degree of polarization in $B$, $V$, $R$, and $I$ bands.}
\label{fig:cdf_km}
\end{figure}

The distributions of linear polarization of active dwarfs in $B$, $V$, $R$, and $I$ bands are shown in Fig. \ref{fig:cdf_km}. The mean value of polarization in $B$ band was found to be $0.16\pm0.01$ per cent, which was more than that observed in other bands. The average values of degree of linear polarization in $V$, $R$, and $I$ bands are found to be $0.080\pm0.006$, $0.056\pm0.004$ and $0.042\pm0.003$ per cent, respectively.

\begin{figure}
\hspace{-5mm}
\includegraphics[width=90mm]{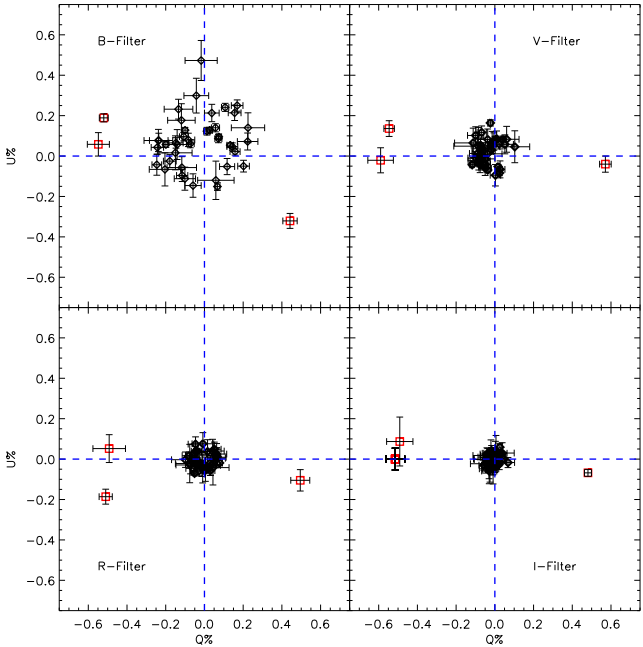}
\caption{Q versus U diagram for the active dwarfs in $B$, $V$, $R$, and $I$ bands.}
\label{fig:qu_stokes}
\end{figure}

\subsection{Polarization due to interstellar medium (ISM)}

\begin{figure}
\hspace{-10mm}
\includegraphics[height=100mm,angle=-90]{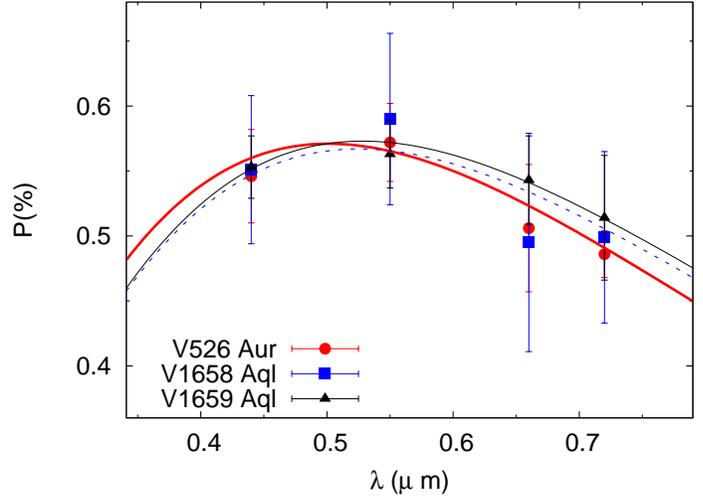}
\caption{The best-fitting Serkowski law for the stars V526 Aur, V1658 Aql and V1659 Aql.}
\label{fig:ism}
\end{figure}

 The majority of stars in our sample are nearby and suffer negligible reddening. Therefore, it is natural to suppose that polarization due to ISM may be negligible. However,  three stars namely  V526 Aur, V1658 Aql and V1659 Aql are located at a distance more than 300 pc and may have suffered considerable reddening indicating that observed polarization may be due to the foreground ISM. Fig. \ref{fig:qu_stokes} shows the plot between normalized Stokes parameters Q and U. The values of normalized Stokes parameters are concentrated near the origin (0,0) of the Q-U graph. However,  the stars V526 Aur, V1658 Aql and V1659 Aql denoted by squares  are located away from the origin indicating external sources of polarization.  Further, in order to explore possibilities of ISM polarization, the average values of polarization are fitted to the standard Serkowski's polarization law (Serkowski et al. 1975):  

\begin{equation}
P_{\lambda} = P_{\rm max} \exp[-K ln^{2}(\lambda_{\rm max}/\lambda)]  
\label{eq:serk}
\end{equation}

\noindent
where $P_{\lambda}$ is the percentage polarization at wavelength $\lambda$ and $P_{\rm max}$ is the peak polarization, occurring at wavelength $\lambda_{\rm max}$. Fig. \ref{fig:ism} shows  the best fit for Serkowski's polarization law  to our BLP data for V526 Aur, V1658 Aql, and V1659 Aql. Adopting the parameter $K = 1.15$ (Serkowski 1973), if the polarization is well represented by the Serkowski relation,  $\sigma_1$ (the unit weight error of the fit) should not be higher than 1.6 due to the weighting scheme; a higher value could be indicative of the presence of intrinsic polarization (see also  Medhi et al. 2010; Eswaraiah et al. 2011). The best-fitting values of $\sigma_{1}$ are found to be 0.893, 0.508, and 0.445 for V526 Aur, V1658 Aql, and V1659 Aql, respectively, which indicate that observed values of linear polarization are due to the ISM.
The best-fitting values of $P_{\rm max}$ and $\lambda_{\rm max}$ are found to be $0.57\pm0.02$, $0.57\pm0.04$ and $0.57\pm0.01$ (in per cent) and $0.50\pm0.05$, $0.53\pm0.07$ and $0.53\pm0.04$ (in $\mu$m) for V526 Aur, V1658 Aql, and V1659 Aql, respectively. The value of $\lambda_{max}$ ~gives  a clue about the origin of polarization. Stars with $\lambda_{\rm max}$ lower than the average value of the ISM ($0.55\pm0.04\mu$m; Serkowski et al. 1975) are  probable candidates to have an intrinsic component of polarization (Orsatti, Vega \& Marraco 1998). The determined values of $\lambda_{\rm max}$ are  very close to the general ISM indicating that these stars have an extrinsic component of polarization.  Using the relation $R_{V} = 5.6 \times \lambda_{\rm max}$ (Whittet \& Van Breda 1978), the total-to-selective extinction ($R_V$) are found to be $2.8\pm0.3$, $3.0\pm0.4$, and $2.9\pm0.2$ for V526 Aur, V1658 Aql, and V1659 Aql, respectively. These values of  $R_V$  are in agreement with the average value of $R_{V}$ (= 3.1) for the Milky Way Galaxy, indicating that the sizes of the dust grains in the direction of these stars are normal.

\subsection{Polarization due to magnetic intensification (MI)}

Polarization due to  MI has been modelled in a number of studies, and the degree of polarization has been assumed to depend linearly on the size of magnetized region of cool stars (e.g. Mullan \& Bell 1976; Calamai \& Degl’Innocenti 1983; Degl’Innocenti 1983; Huovelin \& Saar 1991). Later Saar \& Huovelin (1993) demonstrated that this dependence is non-linear for large magnetic regions, and they developed a model for the wavelength dependence of BLP. The magnetic area relative to the observer  must be asymmetrically distributed for net polarization to result.  Thus polarization can range from zero up to  a maximum value $P_{\rm MI}$ for a given filling factor, depending on the exact arrangement of magnetic regions on the surface. In particular, for a single circular region the degree of polarization will be proportional to a factor of $A$, which approximately depends on the area of the magnetic region ($f$ in per cent).  The maximum expected polarization due to MI is given by Saar \& Huovelin (1993) as 

\begin{figure*}
\hspace{-20mm}
\includegraphics[width=200mm,height=220mm]{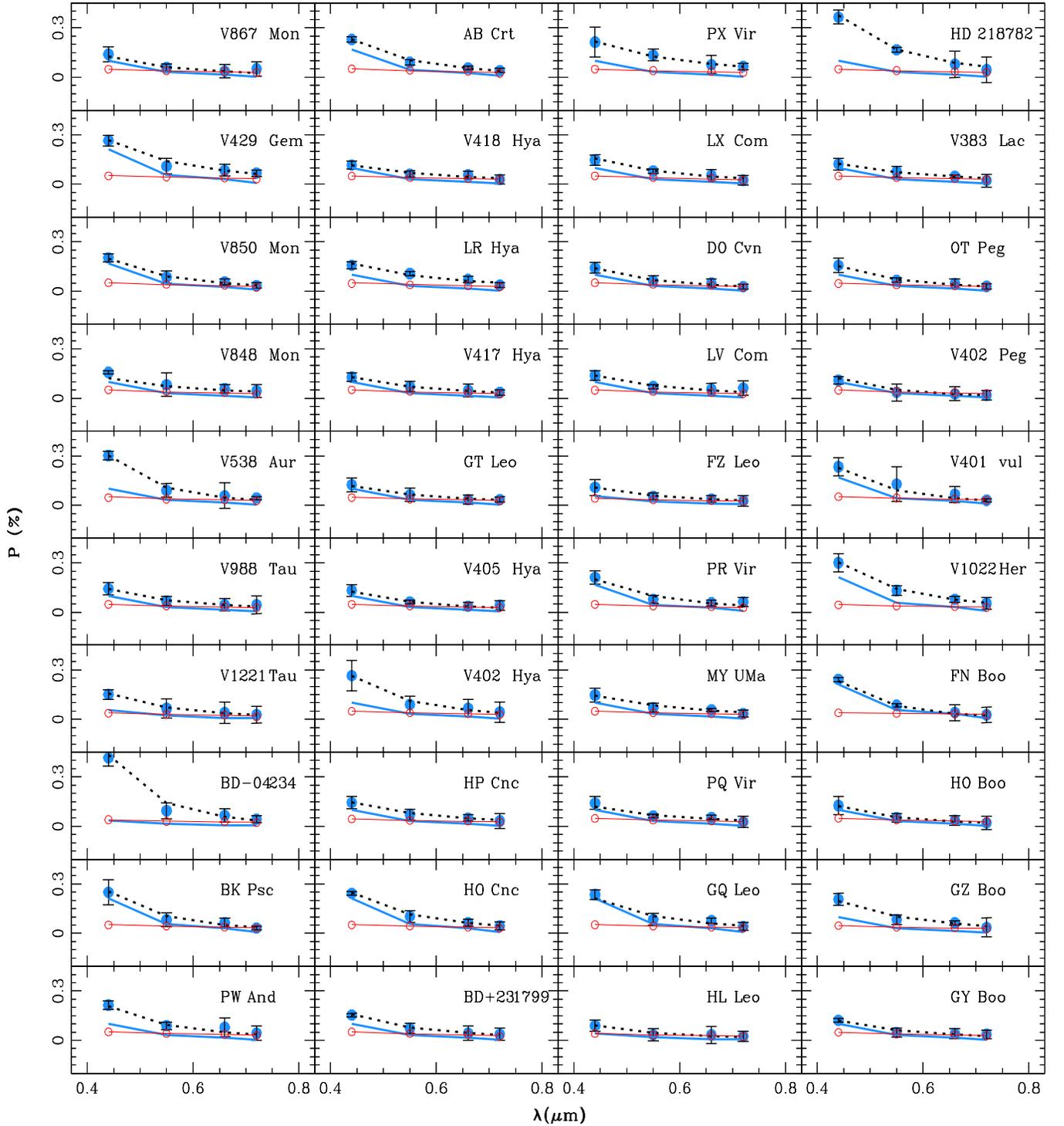}
\caption{Degree of polarization as a function of wavelength for active dwarfs in the sample. The thick continuous lines show the theoretical values of the degree of polarization expected for a Zeeman polarization model. The solid circles with dotted line show the observed values of the degree of polarization. The open circles with continuous line show maximum value of polarization due to scattering.}
\label{fig:wave_depen}
\vspace{-10mm}
\end{figure*}

\begin{equation}
P_{MI} \approx \frac{4}{3\pi}\frac{\xi}{1-\xi}A(f) \Pi(\theta=\frac{\pi}{2})
\end{equation}

\noindent 
where $A(f) \approx -2.128\times10^{-4}+1.076f-4.812f^{2}+9.058f^{3}-6.26f^{4}$, $\theta$ is angle between the magnetic field and the line of sight, $\xi$ is fractional line blanketing and $\Pi$ is  polarized amplitude. Saar \& Huovelin (1993) have calculated a grid of expected degrees of polarization in $U$, $B$, $V$, $R$ and $I$ bands for the stars with temperature ranging  from 4000 to 7000 K, log g from 2.0 to 4.5 and for maximum spot area of 24 per cent.  We have used their results to compare with our observed values of average polarization of active dwarfs. The degree of polarization in $B$, $V$, $R$ and $I$ bands with their theoretical values are shown in Fig.\ref{fig:wave_depen}, where the theoretical values are represented by  continuous thick lines. The observed values of polarization for 12 stars (HP Cnc, FZ Leo, OT Peg, HL Leo, HO Boo, GT Leo, V418 Hya, V401 Vul, BK Psc, V402 Peg, V867 Mon and V988 Tau) are found to be well within 1$\sigma$ level to the theoretical values of the polarization due to MI. However, there are 22 stars where MI polarizations are within 1- 3$\sigma$ from the observed one. The observed values of polarization in six stars (V1221 Tau, HD 218782, V848 Mon, BD+$23^{\circ}~1799$ and V538 Aur) in the sample are found to be above the  3$\sigma$ level from the MI polarizations.

\subsection{Polarization due to scattering}

There are 28 stars, where the observed polarization are well above the $1\sigma$ level from the MI polarization, indicating that polarization in these stars may be due to  additional sources like scattering. 
The extended inhomogeneous envelopes, chromospheres and photospheric spots could be a possible source of scattering induced  BLP in late-type stars.
Stars with chromospheres may also have enough free electrons to cause linear polarization via Thomson scattering (TS), similar to the envelopes of early type stars (Serkowski 1970, Brown \& McLean 1977). However, it has been shown that TS is negligible and Rayleigh scattering is probably a significant contributor to the total BLP. Saar \& Huovelin (1993) derived models which are suitable for the atmospheres of cool stars and to distinguish between scattering and MI as a sources of BLP. These models employed an optimum value for the scattering optical depth ($\tau_{s}) $ of 0.1, whereas for the solar type stars $\tau_{s}$ is generally less than 0.1. A difference in $\tau_{s}$ or in the incident intensity between an asymmetrically distributed region and the surrounding stellar atmosphere is necessary to produce net BLP due to scattering in a star. A horizontally uniform atmospheric layer with uniform incident intensity will yield zero discs integrated BLP for a layer of any $\tau_{s}$. Net scattering induced BLP can be produced by an anisotropic incident light, occurring near the starspots (Finn \& Jefferies 1974). Saar \& Huovelin (1993) derived the following relation for scattering induced BLP:

\begin{equation}
P_{\rm scat}=\frac{3\epsilon}{16(3-\epsilon)}\tau_{s}A_{s}(f)
\label{eq:scat}
\end{equation}

\noindent
where $A_{s}(f)\approx 1.192\times10^{-4}+1.048f-6.945f^{2}+22.46f^{3}-35.92f^{4}+22.55f^{5}$ and $\epsilon$ is the limb darkening coefficient. The maximum scattering polarization for a maximum spot coverage of 18 per cent are computed in $B$, $V$, $R$, and $I$ bands by using the  linear limb darkening coefficients from Claret \& Bloemen (2011) and $\tau_{s}=0.1$. We have used these maximum scattering polarization values to compare with observed polarization values. The maximum scattering polarization for these dwarfs in the sample  are shown in Fig. \ref{fig:wave_depen} with open circles. One can see that maximum polarizations due to the scattering are much  smaller in comparison to the observed polarizations and also less than the polarizations due to MI. This indicates that scattering alone is not responsible for net BLP in these active dwarfs.

\subsection{Polarization due to MI plus scattering}
As we have discussed above the BLP calculated theoretically as arising either due to MI or due to scattering has been 
found to be  less than the observed polarization of majority of stars in the sample. Therefore, in order to see the combined effect of MI and scattering, we have added theoretical maximum values of  polarization due to MI  and  due to scattering. The resultant values are  compared with observed polarization values for  all the stars in the sample. The observed polarization in 29 stars in our sample are found to be well within 1$\sigma$ level to that from the polarizations due to  MI plus scattering. However, there are nine stars (PW And, PX Vir, V402 Peg, V1221 Tau, V402 Hya, HD 218782, GZ Boo, GY Boo and HO Cnc), where combined values of polarization due to the MI and scattering are found to be within 1-3$\sigma$ of the observed values. The observed polarizations in two stars (BD-$04^{\circ}~234$ and V538 Aur) are found to be more than the 3$\sigma$ level of the polarization due to MI plus scattering, indicating supplementary sources of polarization.      

\begin{table}
\centering
{\scriptsize
\caption{Parameters as obtained from the linear regression fit between degree of polarization and various activity parameters along with the correlation coefficient (r) and probability of no correlation (q), where m and c are slope and intercept. }
\label{tab:corr_values}
\begin{tabular}{ccccc}
\hline
Filter      &     $m$               &         $c$          &     $r$   &        $q$      \\ 
\multicolumn{5}{l}{Colour(B-V)}\\
B           &    $0.26\pm0.03$    &   $-0.09\pm0.03$   &   0.55   &     $8.844\times10^{-6}$ \\
V           &    $0.09\pm0.02$    &   $-0.003\pm0.016$    &  0.41   &     0.001            \\
R           &    $0.05\pm0.01$    &   $0.004\pm0.014$    &   0.37   &     0.004             \\
I           &    $0.014\pm0.009$  &   $0.029\pm0.010$    &   0.15   &     0.264            \\
\hline
\multicolumn{5}{l}{Rosby Number($R_{0}$) }   \\
B           &    $-0.12\pm0.02$   &   $0.08\pm0.01$    &  -0.689   &     $8.517\times10^{-9}$ \\
V           &    $-0.051\pm0.009$   &   $0.049\pm0.006$    &  -0.625   &     $4.328\times10^{-7}$ \\
R           &    $-0.042\pm0.007$   &   $0.035\pm0.004$    &  -0.534   &     $3.231\times10^{-5}$               \\
I           &    $-0.010\pm0.006$   &   $0.038\pm0.004$    &  -0.275   &     0.044               \\
\hline
\multicolumn{5}{l}{(R$^{\prime}_{HK}$)}   \\
B           &    $0.094\pm0.044$    &   $0.517\pm0.189$    &   0.479   &     0.001                \\
V           &    $0.062\pm0.014$    &   $0.325\pm0.061$    &   0.459   &     0.002                \\
R           &    $0.033\pm0.010$    &   $0.186\pm0.041$    &   0.492   &     0.001                \\
I           &    $0.024\pm0.007$    &   $0.142\pm0.028$    &   0.357   &     0.017                \\
\hline
\multicolumn{5}{l}{($R_{X} = L_{X}/L_{bol}$)} \\
B           &    $ 0.059\pm0.012$   &   $0.358\pm0.051$    &   0.644   &    $1.369\times10^{-6}$  \\
V           &    $ 0.029\pm0.007$   &   $0.195\pm0.029$    &   0.508   &    0.0003                \\
R           &    $ 0.020\pm0.005$   &   $0.132\pm0.020$    &   0.423   &    0.0034                 \\
I           &    $ 0.008\pm0.004$   &   $0.074\pm0.014$    &   0.149   &    0.3229                \\
\hline

\end{tabular}
}
\end{table}


\section{Linear polarization as a function of different activity parameters}
\label{sec:Correlation}

\begin{figure}
\hspace{-5mm}
\includegraphics[width=90mm]{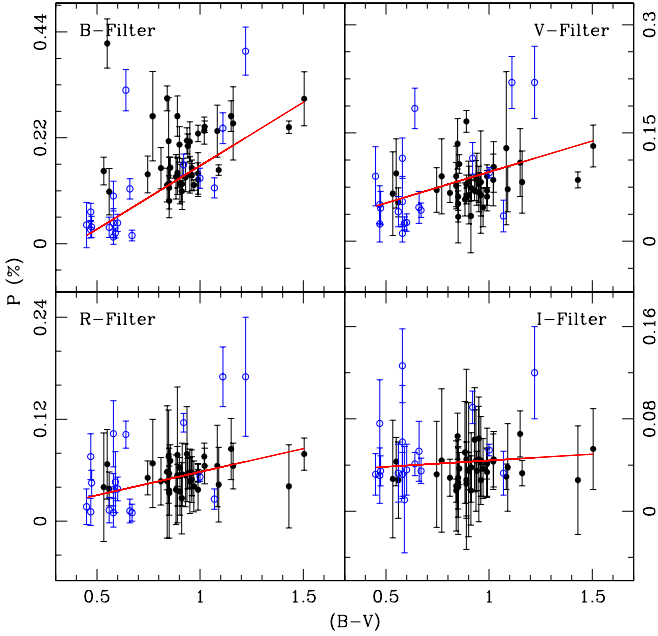}
\caption{Dependence of degree of polarization $P$ on the colour index ($B-V$) in $B$, $V$, $R$, and $I$ bands where solid and open circles represent present data and literature data, respectively and continuous straight line are the linear regression fit.}
\label{fig:colour_fit}
\end{figure}
\subsection{The $(B-V)$ colour}
 The behaviour of degree of linear polarization  $P$ and ($B-V$) colour in $B$, $V$, $R$, and $I$ bands is displayed in  Fig.\ref{fig:colour_fit}. The degree of polarization is found to be increasing towards later spectral type from ($B-V$) = 0.400 to 1.504.  Results of various correlations are given in Table \ref{tab:corr_values}. The correlation coefficients between BLP and ($B-V$) colour are found to be 0.55, 0.41, 0.37, and 0.15 in $B$, $V$, $R$, and $I$ bands,  respectively. The probability of correlation is found to be more than 99 per cent in $B$, $V$, and $R$ bands. However, the correlation in $I$ band is found to be only 70 per cent (see Table \ref{tab:corr_values}).
 The slopes between $P$ and ($B-V$)  are found to be $0.26\pm0.03$, $0.09\pm0.02$, $0.05\pm0.01$ and $0.014\pm0.009$ in the $B$, $V$, $R$ and $I$ bands, respectively.

 \subsection{Rossby Number}
\begin{figure}
\hspace{-5mm}
\includegraphics[width=90mm]{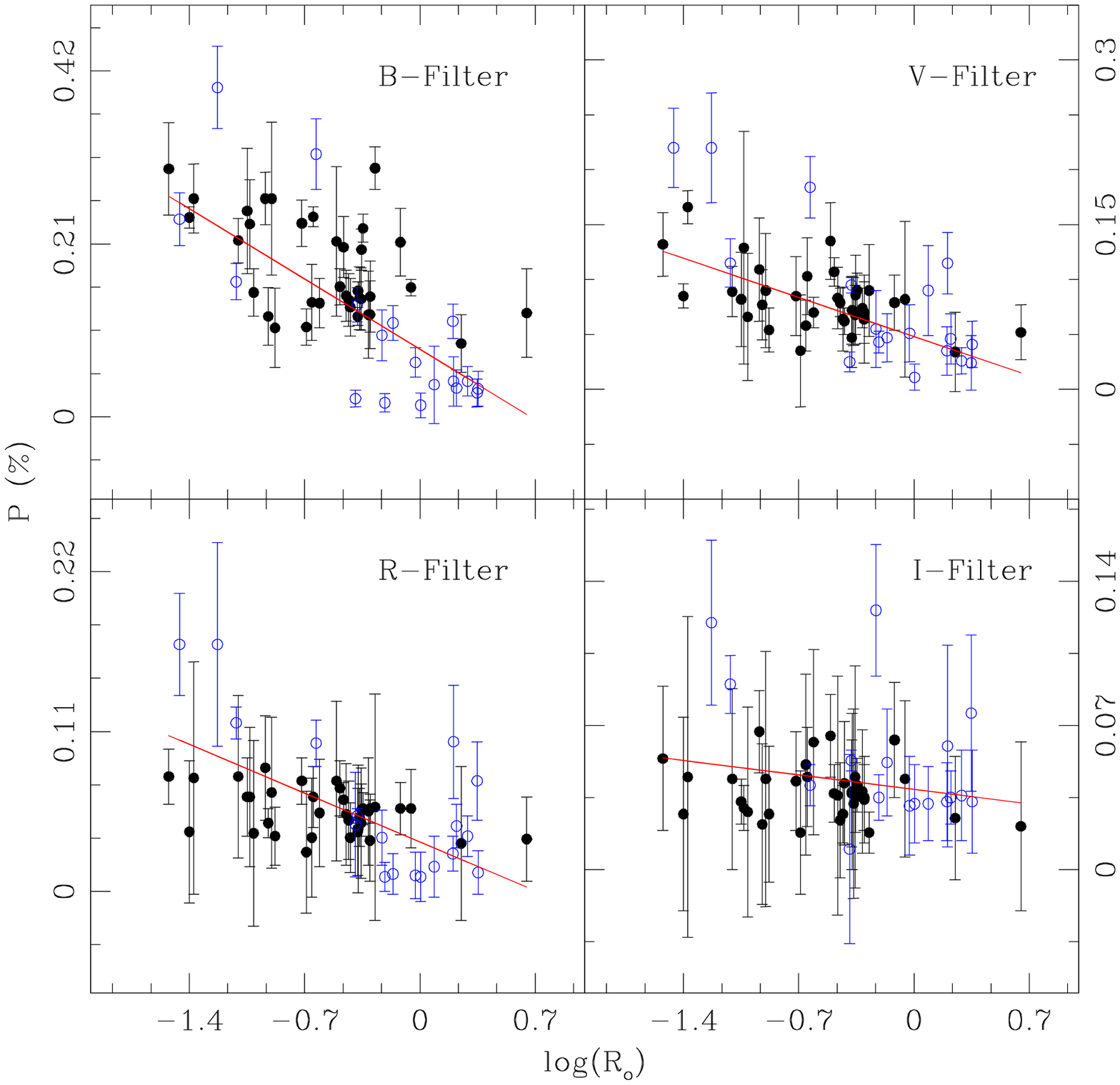}
\caption{Degree of polarization $P$ versus Rossby number $R_{0}$ in $B$, $V$, $R$, and $I$ bands, where symbols are similar to Fig. 5. Straight lines are linear regression fit.}
\label{fig:rosbyno_fit}
\end{figure}

The Rossby number ($R_{o}$) is defined as the ratio of rotational period (P$_{\rm rot}$) and the convective turn over time  ($\tau_{c}$).  The following empirical relation for the convective turn over time in terms of ($B-V$) has been derived by  Noyes et al. (1984) by assuming an intermediate value of ratio of mixing length to scale height (= 1.9).

\begin{equation}
\log\tau_{c} =  \left( \begin{array}{c} 
               1.362-0.166x + 0.025x^{2}- 5.323x^{3}, x > 0\\
                                                           \\
               1.362-0.14x, x < 0 \end{array}\right) 
\end{equation}

\noindent
where $x = 1 - (B-V)$. Making use of this relation we computed values of  $R_o$ and the results are given in Table \ref{tab:object}. The plots between  $P$ and $R_{0}$ in the $B$, $V$, $R$, and $I$ bands are shown in Fig.\ref{fig:rosbyno_fit} . It  is easy to infer from the table that the polarization in $B$ band is strongly dependent on the $R_O$ i.e. the polarization increases with decreasing Rossby number. The dependence of polarization on the $R_O$ is weak towards longer wavelength.   Correlation coefficients for $P$ and log $R_O$ correlation in $B$, $V$, $R$, and $I$ bands have been determined to be of  -0.69, -0.63, -0.53, and -0.28, respectively. The statistical significance of the correlations are found to be more than 99 per cent in each of $B$, $V$, and $R$ bands but for $I$ band no significant correlation was found (see Table  \ref{tab:corr_values}).  The slopes of linear regression fit between these two quantities in the $B$, $V$, $R$, and $I$ are found to be -$0.12\pm0.02$, -$0.051\pm0.009$, -$0.042\pm0.007$, and -$0.010\pm0.006$, respectively.

\subsection{Chromospheric activity index}

The chromospheric activity index ($R^{\prime}_{HK}$) is measured from  chromospheric emission in the cores of the broad photospheric Ca {\sc ii} H \& K absorption lines, normalized to the underlying photospheric spectrum.  $R^{\prime}_{HK}$ is calculated from a band ratio measurement of the Ca {\sc ii} H \& K emission line strength S-index  (Vaughan et al. 1978; Vaughan \& Preston 1980; Duncan et al. 1991) and is given as

\begin{equation}
R^{\prime}_{HK} = \frac{F_{HK}^{\prime}}{\sigma T_{eff}^{4}}
\end{equation}

\noindent
where  $\sigma$ is  Stefan-Boltzman constant and $F_{HK}^{\prime} = 1.6\times10^{6}(10^{-14}S C_{ef}T_{eff}^{4})-F_{0,HK}$ is an excess flux. The conversion factor ($C_{ef}$), basal flux ($F_{0,HK}$), and effective temperature ($T_{eff}$) are  functions of ($B-V$)  and for main sequence stars the empirical relations for these quantities  are given as (Middelkoop 1982; Rutten 1984)  

\begin{equation}
\log(C_{ef}) = 0.25(B-V)^{3}-1.33(B-V)^{2}+0.43(B-V)+0.24
\end{equation}

\begin{eqnarray}
\log(F_{0,HK}) = 1.83-2.76(B-V), {\rm ~~for  ~0.3<(B-V)<0.48} \nonumber \\
= 7.79-2.23(B-V), {\rm ~~for ~0.48<(B-V)<1.25} 
\end{eqnarray}

\begin{equation}
\log(T_{eff}) = 4.04-0.71(B-V)+0.55(B-V)^{2}-0.20(B-V)^{3} 
\end{equation}

\begin{figure}
\hspace{-5mm}
\includegraphics[width=90mm]{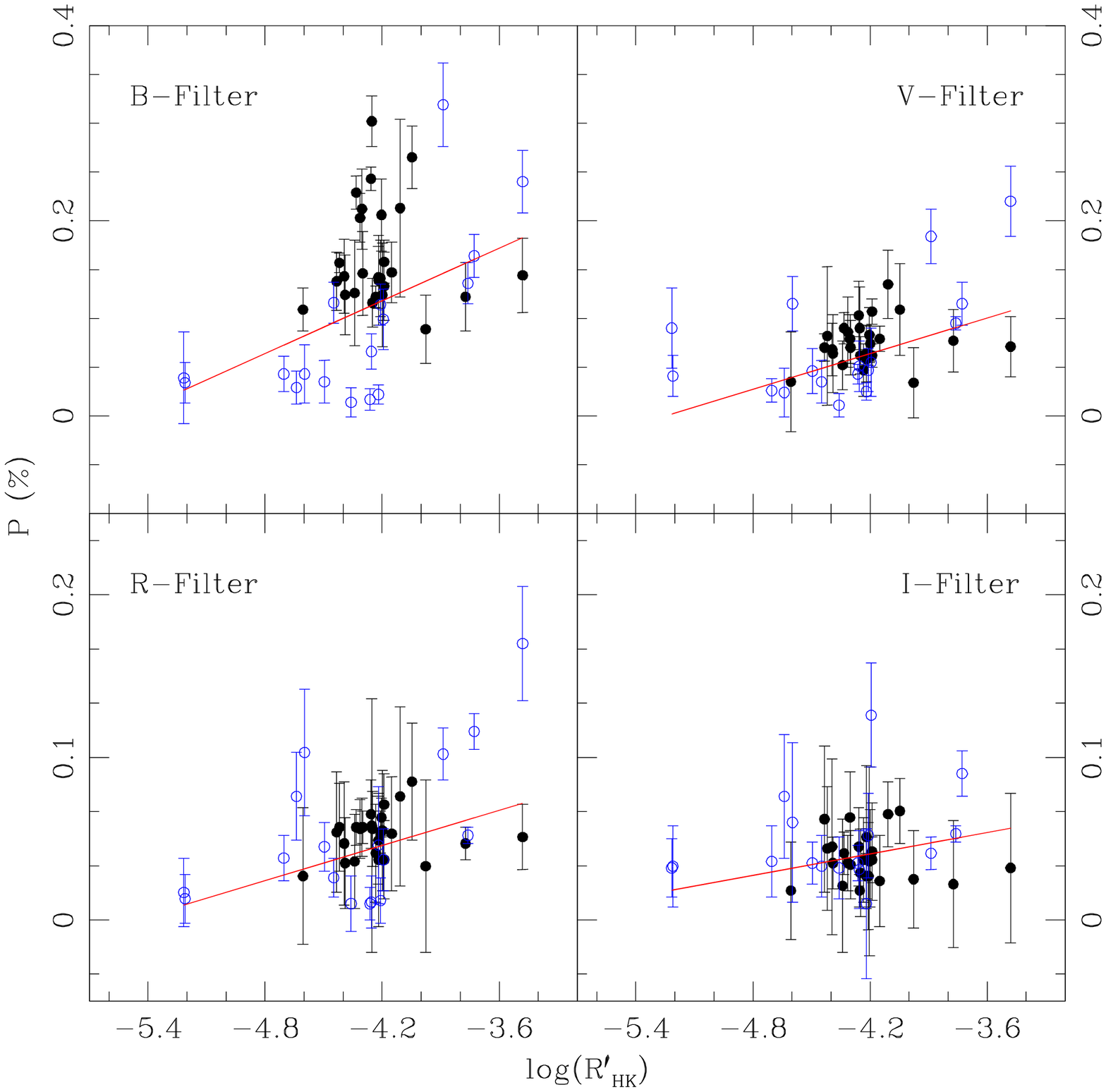}
\caption{Degree of polarization $P$ versus log($R^\prime_{HK}$) in $B$, $V$, $R$, and $I$ bands, where symbols are similar to Fig. 5. Straight lines are the linear regression fit.}
\label{fig:logrhk_fit}
\end{figure}

\noindent
In order to calculate the values  $R^\prime_{HK}$, we have taken the mean of  highest and lowest values S-index from Pace (2013). The resulting logarithmic values of $R^\prime_{HK}$ are given in Table \ref{tab:object}. The dependence of the degree of polarization on $R^{\prime}_{HK}$  in $B$, $V$, $R$, and $I$ bands are shown in Fig.\ref{fig:logrhk_fit}, and results of correlations are given in Table \ref{tab:corr_values}. The solid lines in the figure show the best-fitting straight lines with the slopes as $0.094\pm0.044$, $0.062\pm0.014$, $0.033\pm0.010$, and $0.024\pm0.007$ in the $B$, $V$, $R$, and $I$, respectively. It is observed that the value of slopes have its highest values in the $B$ band and the smallest in the $I$ band. Positive correlations between P and log($R^\prime_{HK}$)  are found with correlation coefficients of  0.479, 0.459, 0.492, and 0.357 in B, V, R, and I bands, respectively.  These correlations are found above  98.3 \% confidence level. 
 
\subsection{Coronal activity index}

\begin{figure}
\hspace{-5mm}
\includegraphics[width=90mm]{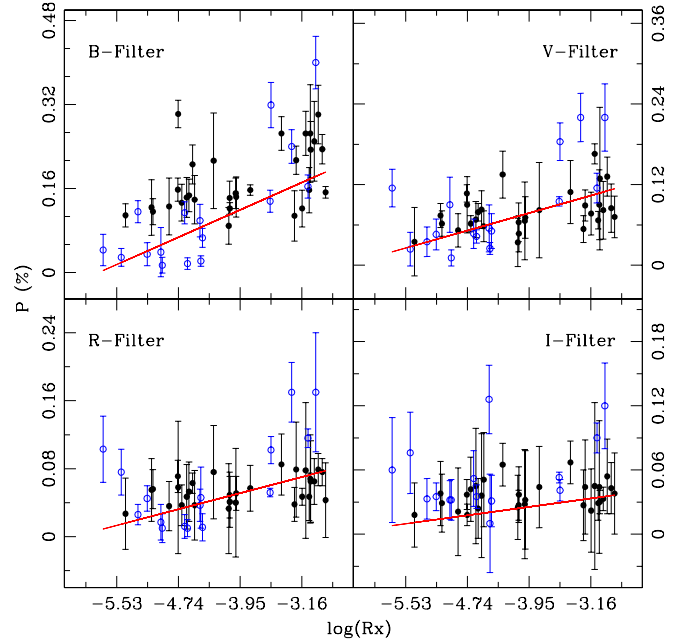}
\caption{Degree of polarization $P$ versus log($R_{X}$) in $B$, $V$, $R$, and $I$ bands, where symbols are similar to Fig 5. Straight lines are the linear regression fit.}
\label{fig:logrx_fit}
\end{figure}

 The coronal activity index ($R_X$) is defined as the ratio of X-ray luminosity ($L_X$) to the bolometric luminosity ($L_{bol}$).  The  $L_X$  of the active dwarfs in the sample are calculated by using relation $L_X = 4\pi D^{2} C_X f_X$, where $D$ is the distance of the star in cm, $f_X$ is the  count rate in counts s$^{-1}$ in the 0.1-2.4 keV energy band, and $C_X = (8.31+5.30HR)\times 10^{-12}$ ~erg ~cm$^{-2}$ ~counts$^{-1}$, is a conversion factor. The  count rate and hardness ratio (HR) were taken  from ROSAT   All Sky Survey (Voges et al. 1999).
The  derived values of $R_X$  are  given in Table \ref{tab:object}.  Fig. \ref{fig:logrx_fit} shows the correlation between degree of polarization and $R_X$ in $B$, $V$, $R$, and $I$ bands, where solid lines are the best-fitting straight lines. Similar to the P versus log ($R^\prime_{HK}$) correlation, a positive correlation was found between P vs log($R_X$). The correlation was also found to be wavelength dependent: strong in $B$ band and weak in $I$ band (see Table  \ref{tab:corr_values}). 
The slopes of linear regression fit  between these two quantities are found to be $0.059\pm0.012$, $0.029\pm0.007$, $0.020\pm0.005$ and $0.008\pm0.004$ in the $B$, $V$, $R$ and $I$ bands, respectively.   Similar to other slopes for above derived correlations, the slopes of  $P$ and log($R_X$) correlation was found to be steeper in $B$-band than that in $V$, $R$, and $I$ bands.

\section{Discussion and Conclusions}
\label{sec:Discussion}
We have carried out the broad-band linear polarimetric observations of a sample of 43 active dwarfs. The degree of polarization in these stars are found to be decreasing towards longer wavelengths with average  values of $0.16\pm0.01$, $0.080\pm0.006$, $0.056\pm0.004$ and $0.042\pm0.003$ per cent in $B$, $V$, $R$, and $I$ photometric bands, respectively. This type of behaviour is similar to that which has also been observed in some other active stars (e.g. Huovelin et al. 1985; Huovelin \& Saar 1988; Yudin \& Evans 2002; Alekseev 2003; Rostopchina et al. 2007; Golovin et al. 2012; Patel et al. 2013). The observed polarizations in these active dwarfs are also comparable to that of  another class of active stars, namely  RS CVn binaries (see Liu \& Tan 1987, Scaltriti et al. 1993). However, the intrinsic polarizations in a few active stars are found to be more than the average values of polarization in  these active dwarfs. Polarization up to 0.6 per cent has been reported in  some other active stars  (e.g. Pfeiffer 1979;  Pandey et al. 2009).  We have investigated the origin of polarization in these stars and have found a variety of results. In three stars, namely  V526 Aur, V1658 Aql and V1659 Aql, the polarization is due to the ISM. However, polarization in 40 stars was found to be intrinsic rather than ISM induced.  
In order to find the origin of polarization in these stars, we have compared the  observed values of polarization with theoretically obtained values of polarization  either due to magnetic intensification or due to scattering.  It appears that the scattering alone is not responsible for the net polarization in  these active dwarfs. In a few active stars, MI is found to be the sole origin of polarization. But in a majority of stars in the sample, the observed polarization is comparable to the sum of polarization due to MI plus scattering. This  indicates that the net polarizations in active dwarfs are due to the combined effect of MI and scattering.  The wavelength dependence of polarization provides another test to distinguish the causes of polarization  in late type active stars whether it is due to magnetic intensification or due to scattering. In order to determine wavelength dependence polarization, we have also fit the power law $P \propto \lambda^{-b}$ to the observed  and theoretical values of polarization. The derived values of power law indices 'b' for the  observed polarization and the theoretical polarization are then compared.  A similar analysis has also been done by Huovelin et al. (1985, 88) for the average values of polarization in $U$, $B$, $V$ and $R$ bands. The values of 'b' are given in Table  \ref{tab:wavelength}, where $b_O$, $b_M$, $b_S$ and $b_{MS}$ are  power law index for observed, MI, scattering, and MI plus scattering polarization, respectively. Only for the star V538 Aur, the values of $b_M$ and $b_O$ were found to be well within $1 \sigma$ level, indicating polarization to be purely magnetic in origin. While comparing  $b_S$ and $b_O$, we found that $b_O$ was well above the $2\sigma$ level from $b_S$  for all the active dwarfs, which further shows that the polarization in these active dwarfs is not due to scattering. However, for most of active dwarfs the values of $b_{MS}$ was comparable with $b_O$ within $1\sigma$ level. This again shows that in a majority of active dwarfs, the linear polarization was due to the sum of polarization due to MI and polarization due to scattering.   There are a few stars, namely BD-$04^{\circ}~234$, V1221 Tau, and V402 Hya, where $b_{O}$ is well above the 3$\sigma$ level than $b_{MS}$, indicating the presence of supplementary sources of linear polarization (see  Pfeiffer 1979; Pandey et al. 2009).  The theoretical  modelling of Saar \& Huovelin (1993) showed that the minimum magnetic field of 2 to 3 kG for K-type stars,  1.3 to 2 kG for G-type stars, and 0.5 to 1.2 kG for F-type stars are required to produce the polarization due to MI. In our sample a majority of stars show the polarization due to MI; therefore, the minimum magnetic field in the sample of stars could be in the range of 0.5 - 3.0 kG.

\begin{table}
\centering
{\scriptsize
\caption{Values of 'b' in the relation $P \propto \lambda^{-b}$ for the polarization due to MI and scattering, and observed polarization.}
\label{tab:wavelength}
\begin{tabular}{clcccc}
\hline
S.No&    Object   & $b_{M}$           & $b_{S}$         & $b_{MS}$        & $b_{O}$ \\
    &             &                   &                 &                 &         \\     
\hline

 1  &     PW And       &  $5.1\pm 0.4$ &$1.0\pm 0.1$ &$2.9\pm 0.2$ &$3.6\pm 0.5$\\
 2  &     BK Psc       &  $5.6\pm 0.5$ &$0.9\pm 0.2$ &$3.8\pm 0.4$ &$3.9\pm 0.5$\\
 3  &     BD-04234     &  $4.0\pm 0.2$ &$1.0\pm 0.2$ &$2.1\pm 0.1$ &$5.0\pm 0.7$\\
 4  &     V1221 Tau    &  $4.3\pm 0.2$ &$1.3\pm 0.4$ &$2.7\pm 0.2$ &$3.5\pm 0.1$\\
 5  &     V988 Tau     &  $5.1\pm 0.4$ &$1.0\pm 0.1$ &$3.0\pm 0.2$ &$2.9\pm 0.3$\\
 6  &     V538 Aur     &  $5.1\pm 0.4$ &$1.1\pm 0.1$ &$3.0\pm 0.2$ &$4.6\pm 0.5$\\
 7  &     V848 Mon     &  $5.1\pm 0.4$ &$1.2\pm 0.3$ &$3.0\pm 0.3$ &$2.6\pm 0.1$\\
 8  &     V850 Mon     &  $5.5\pm 0.6$ &$1.2\pm 0.3$ &$3.6\pm 0.4$ &$3.4\pm 0.2$\\
 9  &     V429 Gem     &  $5.6\pm 0.5$ &$0.9\pm 0.2$ &$3.8\pm 0.4$ &$2.9\pm 0.3$\\
10  &     V867 Mon     &  $5.1\pm 0.4$ &$1.0\pm 0.1$ &$3.0\pm 0.2$ &$3.2\pm 0.8$\\
11  &     BD+231799    &  $5.1\pm 0.4$ &$1.2\pm 0.3$ &$3.0\pm 0.3$ &$3.1\pm 0.2$\\
12  &     HO Cnc       &  $5.6\pm 0.5$ &$0.9\pm 0.2$ &$3.8\pm 0.4$ &$3.4\pm 0.2$\\
13  &     HP Cnc       &  $5.1\pm 0.4$ &$1.0\pm 0.1$ &$3.1\pm 0.2$ &$2.7\pm 0.2$\\
14  &     V402 Hya     &  $5.1\pm 0.4$ &$1.0\pm 0.1$ &$3.0\pm 0.2$ &$3.9\pm 0.1$\\
15  &     V405 Hya     &  $5.1\pm 0.4$ &$1.0\pm 0.1$ &$3.0\pm 0.2$ &$3.1\pm 0.4$\\
16  &     GT Leo       &  $5.1\pm 0.4$ &$1.0\pm 0.1$ &$3.0\pm 0.2$ &$2.7\pm 0.2$\\
17  &     V417 Hya     &  $5.1\pm 0.4$ &$1.2\pm 0.3$ &$3.1\pm 0.2$ &$2.6\pm 0.0$\\
18  &     LR Hya       &  $5.1\pm 0.4$ &$1.1\pm 0.1$ &$3.0\pm 0.2$ &$2.4\pm 0.5$\\
19  &     V418 Hya     &  $5.1\pm 0.4$ &$1.0\pm 0.1$ &$3.0\pm 0.2$ &$2.3\pm 0.5$\\
20  &     AB Crt       &  $5.5\pm 0.6$ &$1.2\pm 0.3$ &$3.7\pm 0.4$ &$3.6\pm 0.2$\\
21  &     HL Leo       &  $4.2\pm 0.2$ &$1.0\pm 0.1$ &$2.2\pm 0.0$ &$3.0\pm 0.6$\\
22  &     GQ Leo       &  $5.6\pm 0.5$ &$0.9\pm 0.1$ &$3.8\pm 0.4$ &$3.1\pm 0.4$\\
23  &     PQ Vir       &  $5.1\pm 0.4$ &$1.0\pm 0.1$ &$3.0\pm 0.2$ &$2.4\pm 0.8$\\
24  &     MY Uma       &  $5.1\pm 0.4$ &$1.0\pm 0.1$ &$3.0\pm 0.2$ &$2.4\pm 0.4$\\
25  &     PR Vir       &  $5.5\pm 0.6$ &$1.0\pm 0.1$ &$3.6\pm 0.4$ &$3.2\pm 0.7$\\
26  &     FZ Leo       &  $4.3\pm 0.2$ &$1.0\pm 0.0$ &$2.4\pm 0.0$ &$2.7\pm 0.3$\\
27  &     LV Com       &  $5.1\pm 0.4$ &$1.0\pm 0.1$ &$3.0\pm 0.2$ &$2.6\pm 0.6$\\
28  &     DO CVn       &  $5.1\pm 0.4$ &$1.2\pm 0.3$ &$3.0\pm 0.3$ &$3.2\pm 0.3$\\
29  &     LX Com       &  $5.1\pm 0.4$ &$1.2\pm 0.3$ &$3.0\pm 0.3$ &$2.9\pm 0.3$\\
30  &     PX Vir       &  $5.1\pm 0.4$ &$1.0\pm 0.1$ &$3.0\pm 0.2$ &$2.5\pm 0.2$\\
31  &     GY Boo       &  $5.1\pm 0.4$ &$1.0\pm 0.1$ &$2.7\pm 0.2$ &$3.0\pm 0.6$\\
32  &     GZ Boo       &  $5.1\pm 0.4$ &$1.0\pm 0.1$ &$3.0\pm 0.2$ &$3.0\pm 0.4$\\
33  &     HO Boo       &  $5.1\pm 0.4$ &$1.0\pm 0.1$ &$3.0\pm 0.2$ &$3.5\pm 0.4$\\
34  &     FN Boo       &  $5.6\pm 0.4$ &$0.6\pm 0.2$ &$3.9\pm 0.4$ &$4.7\pm 0.0$\\
35  &     V1022 Her    &  $5.6\pm 0.4$ &$0.8\pm 0.2$ &$3.8\pm 0.4$ &$3.4\pm 0.2$\\
36  &     V401 Vul     &  $5.5\pm 0.6$ &$0.9\pm 0.1$ &$3.5\pm 0.4$ &$4.1\pm 0.3$\\
37  &     V402 Peg     &  $5.1\pm 0.4$ &$1.0\pm 0.1$ &$3.0\pm 0.2$ &$3.9\pm 0.4$\\
38  &     OT Peg       &  $5.1\pm 0.4$ &$1.0\pm 0.1$ &$3.0\pm 0.2$ &$3.4\pm 0.3$\\
39  &     V383 Lac     &  $5.1\pm 0.4$ &$1.0\pm 0.1$ &$3.0\pm 0.2$ &$2.4\pm 0.3$\\
40  &     HD 218782    &  $5.1\pm 0.4$ &$1.0\pm 0.1$ &$3.0\pm 0.2$ &$3.6\pm 0.1$\\
\hline
\end{tabular}
}
\end{table}

 We have also studied the correlations between  linear polarizations and magnetic activity indicators $R_O$, R$^\prime_{HK}$, and $R_X$. It is found that  the degree of polarization increases with increasing  magnetic activity in late type dwarfs.  Huovelin et al. (1988) also found that the linear polarization is related with  activity indicators, but in a complex way. The slopes of linear regression fit between $P$ and  $R_O$  are found to be steeper than those derived by  Huovelin et al. (1988)  for nine other active dwarfs.  All the  correlations are  found to be similar, which is expected as all the activity indicators are correlated with each other (Noyes et al. 1984, Randich 2000).
These  correlations  are found to be strongest in the $B$ band and also show the largest slope of linear regression fit  in the $B$ band. The correlation becomes weaker towards longer wavelengths. This could be due to  the increase in the number of saturated Zeeman-sensitive absorption lines towards the blue end of the optical spectrum, which in turn produce net linear polarization via the magnetic intensification mechanism and scattering in the optically thin part of stellar atmosphere. The observed polarization is found to be (B-V) colour dependent i.e.  the redder the star, the larger is the observed polarization.
 This is consistent with a trend observed in the $U$ band by Piirola (1977), Tinbergen \& Zwaan (1981), Alekseev (2000) and in $U$, $B$, $V$ bands by Huovelin (1988).
The observed wavelength dependences of linear polarizations and their trend with ($B-V$) support the magnetic intensification mechanism as the dominant contributor to the observed polarization.
       
We conclude that the linear polarization in majority of active dwarfs are due the combined effect of magnetic intensification and scattering, where the degree of polarization is strongly correlated with various activity parameters, especially at the blue end of the optical electro-magnetic spectrum.
 
\section*{Acknowledgements}
This work was done under the Indo-Russian DST-RFBR project reference INT/RUS/RFBR/P-167 (for India) and Grant RFBR Ind\_a 14-02-92694 (for Russia). We thank the referee for reading the manuscript carefully and for  his comments/suggestions. MKP thanks Dr. C. Eswaraiah for useful discussion. SK acknowledges A. Joshi, S. Neha, and B. J. Medhi for their help in observations.

\end{document}